\documentclass[epj,final]{svjour}

\usepackage{amsmath,mathptmx,bm,graphicx,color}

\begin{document}

\title{A sum rule for charged elementary particles}

\author{Gerd Leuchs\inst{1,2}  
\and
Luis L. S\'anchez-Soto\inst{1,2,3}}

\institute{Max-Planck-Institut f\"ur die Physik des Lichts,
G\"{u}nther-Scharowsky-Stra{\ss}e 1, Bau 24, 91058 Erlangen,
Germany
\and
 Institut f\"ur Optik, Information und Photonik,
 Staudtstra{\ss}e 7, 91058 Erlangen, Germany
\and 
Departamento de \'Optica, Facultad de F\'isica, 
Universidad Complutense, 28040 Madrid, Spain}

\date{Received: \today, Revised version: date}

\abstract{There may be a link between the quantum properties of the
  vacuum and the parameters describing the properties of light
  propagation, culminating in a sum over all types of elementary
  particles existing in Nature weighted only by their squared charges
  and independent of their masses. The estimate for that sum is
 of the order of 100.}

\maketitle

\section{Introduction}

The speed of light in vacuum and the impedance of the vacuum for
electromagnetic radiation are experimentally determined parameters,
the value of which has not been deduced so far. The same holds for the
fine structure constant.  Here, {we} use a simple model, borrowed from
the description of dispersion in solid state physics, to attempt
{to establish} a link between classical optics, i.e. Maxwell's
equations, and the relativistic quantum properties of the vacuum.

Maxwell's displacement, $\mathbf{D} = \varepsilon_{0} \mathbf{E} +
\mathbf{P}$, contains a quantity {called the electric
 polarization} of the vacuum. In the SI system, this quantity is
$\varepsilon_{0}$. $\mathbf{P}$ describes the polarization of the
medium, in case we are not dealing with just the vacuum.  Normally,
$\varepsilon_{0}$ is taken as a parameter given by Nature. In the
past, its value has occasionally been adjusted with the availability
of more precise measurement.  Likewise, $1/\mu_{0}$ is the
{magnetization} of the vacuum,\footnote{Initially quantities
  such as the speed of light and the impedance of the vacuum where
  experimentally determined parameters. Then in the SI system in 1948
  the value for $\mu_{0}$ was defined. Later, in 1983 {the} speed
  of light was given a defined value. {As a result,}
  $\varepsilon_{0}$ and the vacuum impedance also had defined
  values. These definitions were made jointly by the institutions in
  charge of standards world wide. The values were defined to be
  compatible with the earlier experimental values within the error
  bars. Currently new SI definitions are being discussed by the same
  institutions with the goal to improve the standards e.g. of the
  kilogram. As a side {effect}, $\mu_{0}$ and $\varepsilon_{0}$
  will be experimentally determined numbers again. For the purpose of
  this paper we, therefore, consider the above constants of classical
  electromagnetism to be experimental numbers, which may tell us
  something about Nature.}  $\mathbf{H} = \mathbf{B}/\mu_{0} -
\mathbf{M}$.  Here we expand on our earlier analysis~\cite{vac1} to
underline {its} relevance for particle physics.  A related
{proposal} linking the quantum vacuum to light propagation was
{obtained} independently by Urban, Couchot and
Sarazin~\cite{UCS2012}.

In the early days of quantum {mechanics}, Weisskopf made the
statement that the positron theory works well provided one ignores any
electric and magnetic polarizability of the vacuum it may
imply~\cite{Weisskopf}.  Looking back, we would reinterpret this
statement as meaning that the polarizability of the virtual
electron-positron pairs in the vacuum must, of course, be already
contained in Maxwell's equations ---otherwise they would not work so
well--- and it would be wrong to account for the same effect a second
time. {However, this implies the} properties of the quantum
vacuum govern the propagation of light and thus govern all of
classical optics. Heitler~\cite{Heitler} likewise mentions that
$\varepsilon_{0}$ may be thought of as the polarizability of the
vacuum associated {with} the electric dipoles induced in the
virtual electron-positron pairs by an external electric field. We now
take this literally and relate the parameters appearing in Maxwell's
equations, $\varepsilon_{0}$ and $1/\mu_{0}$, to the quantum
relativistic properties of the vacuum.  Incidentally, the {term}
Maxwell added to form the Lorentz invariant set of equations,
{he} interpreted as the displacement current of the vacuum. In
our approach, this interpretation comes to life, resulting in a
Lorentz invariant contribution of the quantum vacuum to the
propagation of light.

\section{The model}

The speed of light plays a {multiplicity of  roles in  considerations}
describing different physical quantities: (1) $c_{\mathrm{rel}}$, the
relativistic relation between the mass of a particle and its
{rest} energy
and the limiting speed in the Lorentz transformation; and (2)
$c_{\mathrm{light}}$, the phase velocity of electromagnetic
radiation in vacuum. For the argument below,  we first keep
$c_{\mathrm{rel}}$ and $c_{\mathrm{light}}$ as separate and not
necessarily identical quantities. This obviously means that, for {the}
moment, we relax the requirement for Lorentz invariance of Maxwell's
equations. We derive the speed of light and the impedance of the
vacuum on the basis of the properties of the quantum vacuum treating
it as a dielectric and diamagnetic medium and then compare these
values to the experimentally observed {ones}, thus {restoring} Lorentz
invariance. We will see that the impedance depends on the sum over the
squared charges of the different types of elementary particles, while
the speed of light is independent of this sum. The latter underlines
the general nature of the speed of light. According to present day
knowledge{, the sum reads}
\begin{equation}
\sum_{j}^{{\mathrm{\mathrm{e. \,  p.}}}} \frac{q_{j}^{2}}{e^{2}} = 1 + 1+ 1+ 
 3 \left (3 \frac{1}{9} + 3 \frac{4}{9} \right ) + 1 + \ldots  ? 
\end{equation}
{The sum} has to account for all types of elementary
particle-antiparticle pairs, {known and unknown}.  The known ones
sum up to 9, accounting for electron, muon, tauon, six different
quarks each coming in three colour charges, as well as the charged
$W$-boson.

The model of the vacuum with dielectric and diamagnetic properties
described below is {clearly} oversimplified and a more rigorous model is 
needed. But it will help us getting a first insight into the relation
between the quantum vacuum and optics.

\subsection{The {polarization} of the vacuum}

The vacuum is assumed to consist of virtual particle-antiparticle
pairs treated as extremely short-lived polarizable objects. 
The polarization is a dipole moment density{; therefore,} one has to
calculate the dipole moment induced by an external electric field and
divide by the volume occupied by the pair. We
have two ways of calculating the induced dipole moment.

The first possibility is {to suppose there} is a spring holding a
virtual pair {together}. The spring
constant should be related to the energy needed to excite the virtual
pair to a real pair $\hbar \omega_{0} = 2 m c_{{\mathrm{rel}}}^{2}$.  Since optical
frequencies are a million times smaller than the frequency associated
to the electron-positron energy gap, we are essentially dealing with
the static limit of the driven harmonic oscillator. Note that two
equal harmonically bound masses $m$ correspond to a harmonic
oscillator with only one mass given by the reduced mass
$m_{\mathrm{red}} = m/2$. The corresponding induced electric dipole
moment is
\begin{equation}
  \label{eq:1}
  d = e x = \frac{e^{2}}{m_{\mathrm{red}} \omega_{0}^{2}} \zeta E =  
  \frac{2 e^{2}}{m \omega_{0}^{2}} \zeta  E = 
  \frac{e^{2}  \hbar^{2} \zeta}{2 m^{3} c_{\mathrm{rel}}^{4}} E \,  .
\end{equation}
The factor $\zeta $ accounts for transient creation of the electric
di\-po\-le; i.e. for averaging over the initial transient dynamics leading
to an electric dipole moment smaller than the static limit.  

{The volume} occupied by a single virtual pair can be estimated using the
uncertainty relation and should thus be of the order of the cube of
the Compton wavelength of the electron:
\begin{equation}
  \label{eq:2}
  V = \eta \left ( \frac{\hbar}{m c_{\mathrm{rel}}}\right )^{3} . 
\end{equation}
{We allow for some flexibility
by introducing an additional factor $\eta$, which we expect to be of
order unity}.  The polarization of the electron-positron vacuum is thus
 \begin{equation}
   \label{eq:3}
   P_{0} = \frac{d}{V} = \frac{e^{2} \zeta}{2 c_{\mathrm{rel}}
     \hbar \eta} E \, .
 \end{equation}
{Since the mass drops out,}
different types of elementary particles having the same electric
charge contribute equally to the vacuum polarizability irrespective of
their mass. Hence, to obtain the full vacuum response we have to
sum over all types of  elementary articles, {known and unknown:}
\begin{equation}
  \label{eq:4}
  P_{0} =  \frac{\zeta}{2 c_{\mathrm{rel}} \hbar \eta}
\left ( \sum_{j}^{\mathrm{e. \,  p.}} q_{j}^{2} \right )  E 
 \equiv \varepsilon_{0} E \, .
\end{equation}
As mentioned before, the static limit may be {too large an} estimate for
the induced dipole moment, because the charges have to be accelerated
to this value. The factor $\zeta$ was introduced to account for
this dynamical polarization process. Starting from {zero} 
and averaging over the transiently appearing dipole moment for a time
given by the uncertainty relation for the particle-antiparticle pairs,
one obtains {$\zeta = 1/5$}. Furthermore, we {take the
scale} factor $\eta$  {the same} for all particles.  

{Within this} model one may wonder about a possible frequency
dependence of the vacuum polarization as a result of the resonances at
the rest mass energies.  However, in a real excitation the
conservation of momentum should be fulfilled, prohibiting the
excitation of a virtual pair to a real pair transition in {free space}
with a plane wave. Far away from resonance, the process is
allowed because of the quantum uncertainty of the momentum. Far above
resonance, we would expect the induced dipole moment to decrease as
$1/\omega_{0,j}^{2}$. {In contradistinction}, a converging dipole
wave may excite real pairs in the vacuum~\cite{Narozhny:2004}.

The second alternative of calculating the induced dipole moment is
{to take} the particle and the antiparticle to be free. Accordingly,
the two particles would be accelerated in the external electric field
in opposite directions, but only for the ultra short time during which
we can consider the virtual pairs to exist, which is given by the
relation $\Delta \mathcal{E} \Delta t \ge \hbar$ . The time interval
is thus $\Delta t \simeq \hbar / 2 m_{j} c_{\mathrm{rel}}^{2}$. This
leads to the same expression as in Eq.~(5). Integrating over time in
this free-particle model yields $\zeta = 1/6$, slightly smaller than
the value obtained in the harmonic oscillator model {above.}

\subsection{The magnetization of the vacuum}

Next, we need to {develop} the same procedure for the magnetic
response. The induced magnetic dipole moment $d_{\mathrm{magn}}$ is
given by the current induced in a loop multiplied the loop area:
\begin{equation}
  \label{eq:5}
  d_{\mathrm{magn}} = 2 i A = 2 (q_{j} \nu ) (\pi \varrho^{2}) .
\end{equation}
The factor 2 comes about because the oppositely charged particle and
antiparticle both contribute equal amounts. The frequency at which
the charge goes around the loop is the cyclotron frequency:
\begin{equation}
  \label{eq:6}
  d_{\mathrm{magn}} = \frac{q_{j}^{2}}{m_{j}} \varrho_{j}^{2}  B .
\end{equation}
{The average radius} of the current loop $\varrho_{j}$ {is}  
of the order of the Compton wavelength, with a scale factor $\xi$
{also} taken to be independent of the particle type:
\begin{equation}
  \label{eq:7}
  \varrho_{j}^{2} = \xi  \left ( \frac{\hbar}{m_{j} c_{\mathrm{rel}}}
  \right )^{2} \, .
\end{equation}
Dividing by the volume of the virtual pair [Eq. (3)] we obtain the
vacuum magnetization
\begin{equation}
  \label{eq:8}
  M_{0} = \frac{\xi c_{\mathrm{rel}}}{\eta \hbar} 
  \left ( \sum_{j}^{\mathrm{e. \, p.}} q_{j}^{2} \right )  B \equiv
  \frac{1}{\mu_{0}} B .
\end{equation}
Again the mass drops out and we sum over all types of elementary
particles. We assume the vacuum to be diamagnetic. The
particle-antiparticle pairs will be in singlet states and there will
be no contribution of the total spin of each pair to the magnetization
of the vacuum.~\footnote{If, however, we associate a magnetic
  moment with each particle separately {then} the antiparallel spins
  will lead to parallel yet isotropic magnetic moments in each
  pair. In a way{,} we are making assumptions about the angular momentum
  coupling scheme when neglecting any paramagnetic contribution.}

\subsection{The speed of light and the impedance of the vacuum}

Now,  the {state is set} to relate the speed of light
$c_{\mathrm{light}}$ and the impedance of the vacuum $Z_{0}$ to the
properties of the quantum vacuum. In Maxwell's theory
{$c_{\mathrm{light}}   =1/ \sqrt{\varepsilon_{0}
  \mu_{0}}$} and we can insert the model values for $\varepsilon_{0}$
and $\mu_{0}$ using Eqs.~(5) and (9): 
\begin{equation}
  c_{\mathrm{light}}  =   c_{\mathrm{rel}} \sqrt{\frac{2 \xi}{\zeta}}  \, . 
\end{equation}
Likewise, we find
\begin{equation}
    Z_{0}  =  \frac{\sqrt{2} \eta \hbar}{\sqrt{ \xi\zeta}} 
  \left (     \sum_{j}^{\mathrm{e. \, p.}} q_{j}^{2} \right )^{-1} =
  5811 [\Omega]  \frac{\eta}{\sqrt{ \xi \zeta}}
  \left (  \sum_{j}^{\mathrm{e. \, p.}} \frac{q_{j}^{2}}{e^{2}}
  \right ) ^{-1} . 
\end{equation}
Several things {deserve mentioning}. First of all, if we set the
scale factors to one and the sum over the normalized charges to nine,
we get a speed of light and an impedance which are both off by only a
factor of two. We consider this to be remarkably close {and
  supporting the general applicability of} the model. Secondly, the
speed of light comes out to be independent of how many types of
elementary particles contribute to the {polarization} and
magnetization of the vacuum. This seems to underline the global
characteristics of {$c_{\mathrm{light}}$}.~\footnote{It may be
  worth noting that there seems to be an interesting analogy with a
  completely different quantity, namely the speed of sound in a gas,
  being largely independent of the density.}  We {next} use the
experimental observation that
$c_{\mathrm{light}}=c_{\mathrm{rel}}${; this yields}
\begin{equation}
  \label{eq:9}
  \frac{\zeta}{\xi} = 2 \, .
\end{equation}
{Incidentally,  this results can also be derived through
 requiring that the polarization  and magnetization be the same in all frames}. 
Thus, the expression for the impedance simplifies to
\begin{equation}
  \label{eq:10}
  Z_{0} =  \frac{2 \eta \hbar}{\zeta} 
  \left (     \sum_{j}^{\mathrm{e. \ p.}} q_{j}^{2} \right )^{-1} =
  8218 [\Omega] \frac{\eta}{\zeta}
  \left (  \sum_{j}^{\mathrm{e. \, p.}} \frac{q_{j}^{2}}{e^{2}}
  \right ) ^{-1} . 
\end{equation}

\subsection{The fine structure constant}

The fine structure {constant} $\alpha$ relates to the strength of the
coupling between the electromagnetic field and matter and is given by 
\begin{equation}
  \label{eq:alphadef}
  \alpha = \frac{e^{2}}{4 \pi \varepsilon_{0} \hbar c} 
\end{equation}
This is the zero-energy value. There is experimental evidence for an
increase of $\alpha$ toward higher energies~\cite{Levine:1997},
referred to as the running fine structure constant. This is
{usually ascribed to} the renormalization of the electric charge.

In  quantum electrodynamics, one could modify $\varepsilon_{0}$
instead of renormalizing the electric charge. Our model suggests doing
exactly this by relating  $\varepsilon_{0}$ to the sum over the
different types of elementary particles
\begin{equation}
  \label{eq:e0de}
  \varepsilon_{0} =    \frac{\zeta}{2 c \hbar \eta }
\left ( \sum_{j}^{\mathrm{e. \,  p.}} q_{j}^{2} \right ) \, ,
\end{equation}
which results in the {following expression} for the zero-energy value of the
fine structure constant:
\begin{equation}
  \label{eq:alphadef2}
  \alpha_{0} = \frac{\eta}{2 \pi \zeta} \left (
\sum_{j}^{\mathrm{e. \,  p.}} \frac{q_{j}^{2}}{e^{2}} \right
  )^{-1}  \, . 
\end{equation}
The model prediction for the value of $\alpha_{\mathcal{E}}$ at higher
energies is obtained by omitting those particle-antiparticle pairs
having a rest mass energy lower than $\mathcal{E}$. This allows for an
alternative route to estimate the sum in Eq.~(1),

\subsection{Dispersion}

One consequence of this model is that as soon as the light frequency
increases beyond the gap frequency for one particular
particle-antiparticle pair, the contribution of this pair to the sum
will decrease,~\footnote{{When the} frequency of the electromagnetic wave
  $\omega$ increases beyond the rest mass energy of one type of
  particle-antiparticle pair, then the contribution of this type of
  particles to the sum in Eq.~(1) drops to zero at a rate proportional
  to the inverse of the frequency squared for the electric
  polarizability and at a rate proportional to the inverse of the
  frequency for the magnetic polarizability. The electric dipole
  moment is induced with some delay owing to the inertia of the
  particle mass, while the magnetic dipole moment is induced
  instantaneously for a point charge. Thus, one would not expect any
  frequency dependence of the magnetic polarizability of point
  charges. However, the position uncertainty of the order of the
  Compton wavelength leads to a reduced current at frequencies higher
  than the resonance and thus to the inverse frequency dependence.}
both for $\varepsilon_{0}$ and $1/\mu_{0}$. Within this model the sum
cancels out when calculating the speed of light, predicting a
frequency independent speed of light equal to $c_{\mathrm{rel}}$. The
vacuum impedance however will be affected starting at gamma ray
frequencies above 1 MeV. It would certainly be interesting
{to investigate this prediction} for a modification of the quantum
vacuum.~\footnote{{If} the statement about the different
  frequency dependences in the electric and in the magnetic case in
  footnote 4 is correct, there might be deviations to the speed of
  light in the vicinity of the rest mass energies.}

\section{Discussion}

There are two independent comparisons between the model and
experimental {values} both leading to a prediction for the
{sum} over all charged particles, known and unknown ones.

\subsection{The number of charged elementary particles as derived from
  the impedance of the vacuum}

We can now set the model result in Eq.~(13) for the vacuum impedance
equal to the empirical value $Z_{0} = 376.7$ $\Omega$. The remarkable
result is that this provides information about {future
  additional}  types of charged particle-antiparticle pairs:
\begin{equation}
  \label{eq:11}
  \sum_{j}^{\mathrm{e. \, p.}} \frac{q_{j}^{2}}{e^{2}}  = \frac{2 \eta
    \hbar}{e^{2} Z_{0} \zeta} = 21.82 \frac{\eta}{\zeta} \, . 
\end{equation}
There are still two unknowns: the sum and $\eta$. Without any further
information we use the initial assumption in Sec. 2.1 that $\eta \simeq
1 \pm \delta \eta$:
\begin{equation}
  \label{eq:malsu}
  \sum_{j}^{\mathrm{e. \,  p.}}  \frac{q_{j}^{2}}{e^{2}}  
\simeq 109  (1 \pm  \delta \eta  )   \, .
\end{equation}
Here, $\delta \eta$ accounts for the uncertainty in the value 
of $\eta$. 

\subsection{The number of charged elementary particles as derived from
  the energy dependence of the fine structure constant}

A second independent estimate for the sum over all particles involves
the fine structure constant and its established variation with
energy. The fine structure constante is $1/137.04$ at low energies and
reduces to $1/(128.5 \pm 2.5)$ at 58~GeV.  At this energy we are
beyond the rest mass energies of most of the well-known
particle-antiparticle pairs (except for the top quarks and the W
bosons). So, {omitting the particle types with $m_{j} c^{2} < 58$
  GeV the sum would reduce by 20/3 = 6.67}, increasing $\alpha$
correspondingly. Based on our model, and with the experimental values
for $\alpha_{0}$ and $\alpha_{\mathrm{58 \, GeV}}$, we find a second
independent way to determine the sum:
\begin{eqnarray}
  \alpha_{0}^{-1} & = & 137.04  =   \mathrm{constant}  \sum_{j}^{\mathrm{e. \,
      p.}} \frac{q_{j}^{2}}{e^{2}} \, , \nonumber \\
  & & \\
  \alpha_{\mathrm{58  \, GeV}}^{-1} & = & 128.5 \pm 2.5  =   
  \mathrm{constant}  \sum_{j}^{\mathrm{e. \,  p.> 58 \, GeV} } 
  \frac{q_{j}^{2}}{e^{2}} \, . \nonumber 
\end{eqnarray}
 In the calculation, we {can} take into
account that the contribution of one particle type does not fall off
abruptly but proportional to $1/\omega$, {so each term in the sum
  is replaced with}
\begin{equation}
  \label{eq:17}
\frac{q_j^2}{e^2} \mapsto \frac{q_j^2}{e^2}  \times 
 \left \{
\begin{array}{ll}
1 & \qquad \hbar \omega \le m_{j} c^{2} \\
 & \\
\displaystyle
\left ( \frac{m_j c^2}{\hbar \omega} \right ) ^{2}&  \qquad \hbar \omega > m_{j} c^{2} 
\end{array}
\right  .
\end{equation}
This results in a reduction of the sum for 58 GeV by 6.5 instead of  6.67. 
{Consequently, }
\begin{equation}
  \frac{\alpha_{0}^{-1}}{\alpha_{\mathrm{58  \, GeV}}^{-1}} =  
  \frac{\sum_{\mathrm{all}}}{\sum_{\mathrm{> 58 \, GeV} }} = 
  \frac{\sum_{\mathrm{all}}}{\sum_{\mathrm{all}} - 6.5} =
  \frac{137.04}{128.05 \pm 2.5} \, ,
\end{equation}
so finally 
\begin{equation}
 \sum_{\mathrm{all}} = 104  \left \{ \begin{array}{l}  
+ 43 \\
- 24 
\end{array}
\right . 
\end{equation}

\subsection{Comparison}
 
The two different results obtained so far agree quite well. It seems
that the approach in 3.2 is less ambiguous than the one in 3.1. On
could use Eq.~(21) to reduce this ambiguity in 3.1. Using $\zeta =
0.2$ and imposing the result in (21) would lead to $\eta = 1.05 \pm
0.31$. This number is very close to the one assumed in section
3.1. Consequently, when using $\zeta = 0.2$, $\xi = 0.1$ and $\eta
=1.05$, the two approaches are both compatible, the sum being of order
of 100. This would predict many still undiscovered charged elementary
particles with rest mass energies above 58~GeV.

\section{Conclusions}

Our model is a most simple one and the quantitative results, namely
the sum over the different types of elementary particles have thus to
be taken with caution. One feature of this model is that it {relates
  the number of charged elementary particles to low-energy} properties
of the electromagnetic field, such as the vacuum impedance and the
fine structure constant. The zero energy value of the fine structure
constant, or equivalently the vacuum permittivity, has so far been a
purely experimental number. {As to} the speed of light, the value
predicted by the model is determined by the relative properties of the
electric and magnetic interaction of light with the quantum vacuum and
is independent of the number of elementary particles, a remarkable
property underlining the general character of the speed of light.

Thus, the purpose of the simple model is to point at the intimate
relationship between the properties of the quantum vacuum and the
constants in Maxwell's equations. {Indeed, from this picture, the
  vacuum can be understood as an effective
  me\-dium~\cite{Dicke:1957}. Furthermore, we have devised two
  independent ways of checking the model predictions against the
  experimental values}. We hope that this result will stimulate more
rigorous quantum field theoretical calculations.

\section*{Acknowledgments}

  It is a pleasure acknowledging helpful discussions with Joseph~H. 
  Eberly, Holger Gies, Nicolai B. Narozhny, Wolfgang P. Schleich,
  Michael Thies, Joe Marcel Urban, Fran\c{c}ois Couchot and Xavier
  Sarazin, as well as with the participants of the 500th Heraeus
  Seminar on "Highlights of Quantum Optics" at the Physik-Zentrum in
  Bad Honnef, May 6-11, 2012.


\begin{thebibliography}{99}

\bibitem{vac1}
G. Leuchs, A. S. Villar, L. L. S\'anchez-Soto,
Appl. Phys. B \textbf{100}, 9 (2010).

\bibitem{UCS2012}
M. Urban, F. Couchot, X. Sarazin: quantum-ph: 1106.3996, 1111.1847

\bibitem{Weisskopf}
V. Weisskopf, in: J. Schwinger, 'Selected papers on
  Quantum Electrodynamics', p.92-128 (Dover, New York, 1958). 

\bibitem{Heitler}
W. Heitler, "The quantum theory of
  radiation", p.113, 3rd edition, (Oxford University Press, Oxford, 1954).

\bibitem{Levine:1997}
I. Levine \emph{et al}, Phys. Rev. Lett. \textbf{78}, 424 (1997).

\bibitem{Narozhny:2004}
N. B. Narozhny, S. S. Bulanov, V. D. Mur, and V. S. Popov, 
Phys. Lett. A \textbf{330},  1 (2004).

\bibitem{Dicke:1957}
R. H. Dicke Rev. Mod. Phys. \textbf{29}, 363 (1957)
\end{thebibliography}
\end{document}